# Self-organised complex aerial displays of thousands of starlings: a model

H. Hildenbrandt[a], C. Carere[b,c]., C-K. Hemelrijk[a*]

[a]*Theoretical Biology, Centre for Ecological and Evolutionary Studies, University of Groningen. P.O. Box 14, 9750 AA Haren, The Netherlands,*

[b]*CNR - INFM, Dipartimento di Fisica, Universita' di Roma La Sapienza, P.le A. Moro 2, 00185 Roma, Italy,*

[c]*Section of Behavioural Neurosciences, Department of Cell Biology and Neurosciences, Istituto Superiore di Sanità, Viale Regina Elena 299, 00161 Rome, Italy.*

[*] Corresponding author: *E-mail address:* C.K.Hemelrijk@rug.nl. Tel. 0031-50-3638084, Fax. 0031-50-3633400.

**Abstract**

Aerial displays of starlings (*Sturnus vulgaris* ) at their communal roosts are complex: thousands of individuals form multiple flocks which are continually changing shape and density, while splitting and merging. To understand these complex displays both empirical data and models are needed. Whereas detailed empirical data were recently collected through video recordings and position measurements by stereo photography of flocks of thousands of starlings, there are as yet no models that generate these complex patterns. Numerous computer models in biology, however, suggest that patterns of single groups of moving animals may emerge by self-organisation from movement and local coordination (through attraction, alignment and avoidance of collision). In this paper, we investigated whether this approach can be extended to generate patterns resembling these aerial displays of starlings. We show in a model that to generate many of the patterns measured empirically in real starlings we have to extend the usual rules of local coordination with specifics of starling behaviour, mainly 1) their aerial locomotion, 2) a low and constant number of interaction-partners and 3) preferential movement above a 'roosting area'. Our model can be used as a tool for the study of these displays, because it provides new integrative hypotheses about the mechanisms underlying these displays and of swarming patterns in biological systems in general.

**Introduction**

Moving animal groups (including those of humans) often exhibit complex patterns of coordination. One of the most complex patterns is shown by flocks of tens of thousands of European Starlings (*Sturnus vulgaris*) over the roost before nesting at dusk. Because the coordination during their wheeling and turning, splitting and merging and the changes of flock shape and density are truly amazing, this performance has been attributed to 'thought transference' (Selous, 1931). To understand how these patterns are generated is not only important in itself, but also for obtaining insight how starlings –despite their large numbers- are able to prevent collisions even when escaping a predator (Carere et al., 2009; Feare, 1984). This may bear interest in relation to traffic control (Helbing and Huberman, 1998; Lubashevsky et al., 2003). To increase our understanding of collective patterns of motion, a combination of empirical and theoretical studies is needed. The first 3D reconstruction of huge swarms comprising thousands of starlings have been made only recently (Ballerini, et al., 2008a; Ballerini et al., 2008b; Cavagna et al., 2008) due to the great difficulty of collecting empirical data on individual positions in a flock (Budgey, 1998; Major and Dill, 1978; Pomeroy and Heppner, 1992). Furthermore, extensive video recordings of the aerial displays have been made (Carere et al., 2009). Modelling studies of starling displays however are still lacking.

So far biologically-oriented models of swarming have shown that stable swarming patterns may arise by self-organisation from rules for movement and interactions with others close by (Buhl et al., 2006; Couzin et al., 2005; Helbing et al., 2005; Hemelrijk, 2002; Parrish and Edelstein-Keshet, 1999). Here we develop the first model that apart from rules for movement and local coordination also incorporates a simple representation of aerodynamics during aerial locomotion, banking during turns and staying above the roost. We compare its emergent patterns to the recent empirical studies of displays of starlings in Rome (Ballerini et al., 2008b) as regards shape, internal structure and the general dynamics of the movements of the flocks and their splitting and merging.

In designing our new model, we have followed the approach of complexity science, in which it is argued that complex patterns of collective behaviour are not due to high cognitive abilities, but to interactions of individuals with each other and with their environment, and to the way in which these interactions are shaped by the structure of their body (embodiment) (Pfeifer and Scheier, 1999). Accordingly, in developing our model of starling displays we started from the rules of local coordination among individuals, and added strongly simplified aerodynamics of flying behaviour as well as a tendency to stay in a cylindrical area of approximately the size of the flight area above the roosting site at Termini in Rome (Ballerini et al., 2008a; Ballerini et al., 2008b; Carere et al., 2009). As the starting point of our simulation, we used our former model of schools of fish (Hemelrijk and Hildenbrandt, 2008), because it solved a number of problems for which other related models have been criticized (Parrish and Viscido, 2005). They were criticized for being based on fixed speed, global perception and small group sizes. We corrected these shortcomings in our recently published fish model (Hemelrijk and Hildenbrandt, 2008); in this model, individuals can change their speed, they perceive only those others that are closest to them, because the distance of perception is reduced when they are surrounded by others more densely and they travel in small as well as very large schools of some thousands of fish in real time (Hemelrijk and Hildenbrandt, 2008).

In short we developed a model of the self-organisation of starling-like displays at the roost. We will show that patterns of our model resemble starling-like displays qualitatively in their shape, splitting and merging, movement trajectories, positioning of individuals during turning, dynamics over time, and the occurrence of banking of the entire flock in the direction of banking. Furthermore, results of the model were nonsignificantly different from the empirical data for more than half (10 out of the 18) of the statistical test we performed. This means that we captured a number of essential characteristics of these displays. We will indicate how our model may be useful for future studies of flocks.

**Methods**

**The Model**

In developing our model for starlings we have combined components from our model for schooling by fish (Hemelrijk and Hildenbrandt, 2008) with traits that are characteristic of starling behaviour at the roost. The model is developed in realistic units of forces and speed and its parameters are tuned to realistic values wherever available (Tab. 1).

Each individual is characterised by its mass, $m$, its speed, $v$, and its location, $r$. Further, the orientation in space of each individual is given by its local coordinate system ($e_x$, $e_y$, $e_z$). Like in the model by Reynolds (Reynolds, 1987), its orientation is indicated by its forward direction, $e_x$, its sideward direction, $e_y$, and its upward direction, $e_z$, which it changes by rotating around these three principal axes, $e_x$, $e_y$ and $e_z$ (*roll*, *pitch* and *yaw*) (Fig. 1). The behaviour of each individual is based on its cruise speed, its social environment (i.e. the position and movement direction of its nearby neighbours),

Self-organised starling displays, 6

its attraction to the roost and the simplified physics of aerodynamic flight which includes banking while turning.

Like in our earlier model, an individual $i$ tends to move at cruise speed $v_0$ and a force, $f_{\tau_i}$ (Equ. 1), brings it back to it after it has deviated from it to avoid collision or to catch up with others (Hemelrijk and Hildenbrandt, 2008).

$$f_{\tau_i} = \frac{m}{\tau}(v_0 - v_i) \cdot e_{x_i}, \qquad \text{Speed control (1)}$$

where $\tau$ represents the relaxation time, $m$ is the mass of the individual $i$ and $v_0$ its cruise speed, ($m$ and $v_0$ are the same for all individuals), $v_i$ is its velocity, and $e_{x_i}$ its forward direction.

Similar to other models (Couzin et al., 2005; Couzin et al., 2002; Hemelrijk and Hildenbrandt, 2008; Reynolds, 1987) individuals coordinate with each other through 'social forces' $F_{Social}$ that are Newtonian (Equ. 4-8): while they avoid collision with

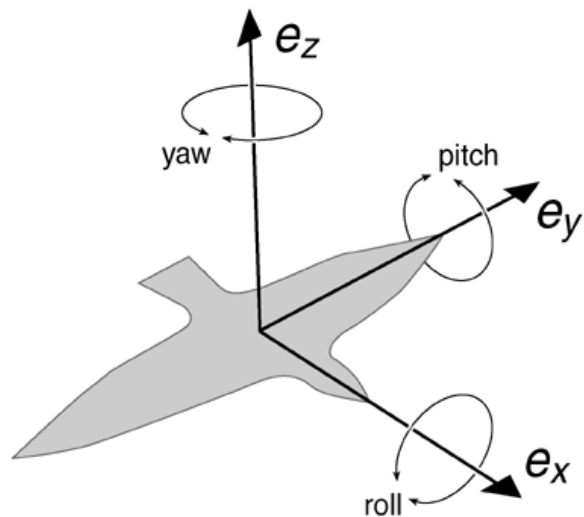

**Figure 1.** Local coordinate system. $e_x$ is the bird's forward direction, $e_y$, its sideward direction and $e_z$, its upward direction. It can change these by rotating around these each other (separation), individuals at the same time group by being attracted to others (cohesion) and by trying to move in the same direction (alignment). We have adjusted these rules in a number of ways to the behaviour of starlings.

For instance, starlings interact only with a fixed number of 6-7 nearest neighbours, the so-called 'topological range', (Ballerini et al., 2008a; Ballerini et al., 2008b), independent of their distances. This differs from a metric range of interaction in which individuals interact with all others in a fixed metric distance as has been the usual implementation in models of swarming (Couzin et al., 2002; Hemelrijk and Kunz, 2005; Huth and Wissel, 1994; Romey, 1995). To represent topological interaction, each individual $i$ in the model adapts its metric interaction range, $R_i(t)$ (Hemelrijk and Hildenbrandt, 2008) such that individuals attempt to interact only with a constant number of their closest neighbours (Equ. 2,3).

$$R_i(t + \Delta u) = (1 - s) R_i(t) + s \left( R_{max} - R_{max} \frac{|N_i|}{n_c} \right) \qquad \text{Adaptive interaction range (2)}$$

$$N_i \stackrel{def}{=} \{j \in N; d_{ij} \leq R_i\} \qquad \text{Neighbourhood of an individual (3)}$$

where $\Delta u$ is the time to react, $s$ is the interpolation factor, $R_{max}$ is the maximal interaction range, $N_i$ is the neighbourhood of individual $i$, i.e. the set of neighbours of an individual $i$ which is composed of $|N_i|$ neighbours, $n_c$ is the fixed number of topological interaction partners it strives to have and $d_{ij}$ is the distance between individual $i$ and $j$. Thus, the radius of interaction at the next step in reaction-time, $R_i(t+\Delta u)$, is increased if the number of interaction partners $|N_i|$ is smaller than the targeted fixed number $n_c$, it is decreased if it is larger than the fixed value and it remains as before if $|N_i|$ equals $n_c$ (Fig. 2). Here $R_i$ can not increase beyond $R_{max}$ and $s$, the interpolation factor, determines the step-size of the changes. As a result, $|N_i|$ fluctuates around $n_c$ with a variance that is controlled by the interpolation factor $s$.

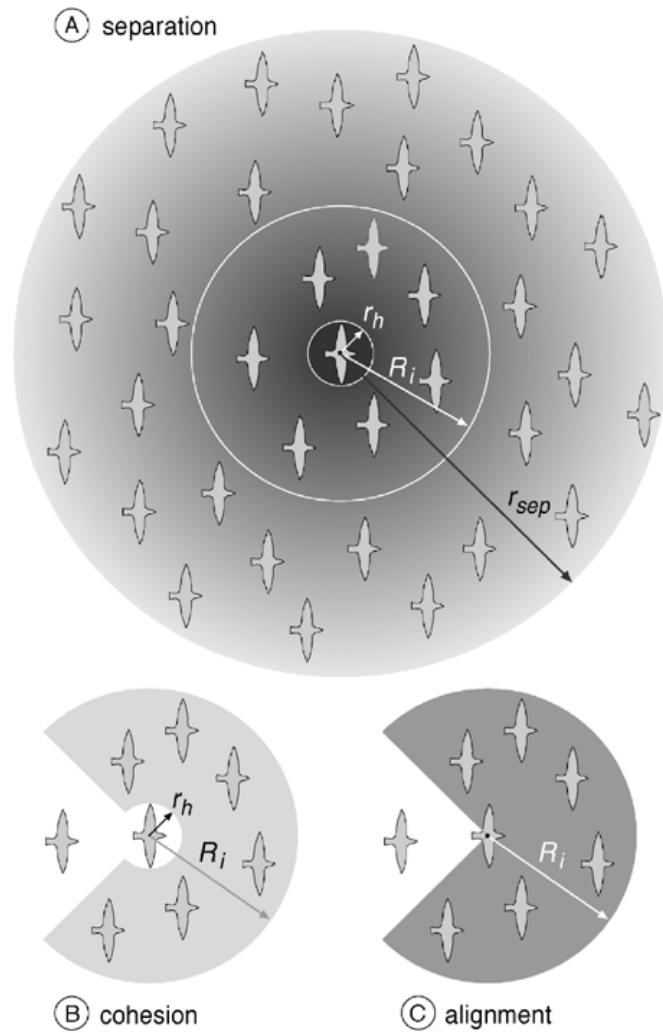

**Figure 2.** Social interaction ranges for separation (A), cohesion (b) and alignment (C). Note that the values of the different radii in the figure are not to scale with the default values in Tab. 1.

As to separation, individual $i$ perceive a force $f_{s_i}$ to move in the opposite direction of the average direction of the locations of $N_i$ others in its neighbourhood (Fig. 2). We omitted the blind angle at the back that is usual in these models (Equ. 4 and see Parameterization & Experiments). Further, because in empirical data individuals avoid coming too close to each other, and they are surrounded by a 'hard sphere' (Ballerini et al., 2008b) which represents the volume where an individual's body is located and its wings are flapping, we represented this hard sphere (with diameter equal to the bird's wingspan) in the model (Fig. 2). In this hard sphere with radius $r_h$ separation is maximal (Equ. 4) and cohesion is absent (Equ. 5). Outside the hard sphere, but inside the radius of separation $r_{sep}$ the separation force declines following a halved Gaussian, $g(x)$, with $\sigma$ the standard deviation of the Gaussian set such that at the border of the separation zone the force is almost zero, $g(r_{sep}) = 0.01$.

$$f_{s_i} = -\frac{w_s}{|N_i|} \sum_{i \in N_i} g(d_{ij}) d_{ij} \;; \qquad g(x) = \begin{cases} 1 & ; x \leq r_h \\ \exp\left(-\frac{(x - r_h)^2}{\sigma^2}\right) & ; x > r_h \end{cases} \qquad \text{Separation (4)}$$

Here, $|N_i|$ is the number of individuals in the neighbourhood of interaction and the distance $d_{ij}$ from individual $i$ to individual $j$ is given by $\|p_j - p_i\|$, where $p_i$ gives the position of an individual $i$. Further, the direction from individual $i$ to individual $j$ is specified by the unit vector $d_{ij}=(p_j - p_i)/\|p_j - p_i\|$ and $w_s$ is the fixed weighting factor for separation (Tab. 1). Note that if a flock is very sparse, a few of the interaction partners may be located outside the range of separation. Therefore, these individuals will only be subject to alignment and cohesion. Further, note that interactions to avoid others are not entirely topological, because distance matters also, as it determines the magnitude of the force (it is larger if others are closer and particularly large in the hard sphere).

As to cohesion, individual $i$ is attracted by a force $f_{c_i}$ to the direction of the centre of mass the centre of mass (i.e. the average $x$, $y$, $z$ position) of the group of $N^*_i$ individuals located in its topological interaction neighbourhood. Note that in this case we exclude those individuals of the set of $N_i$ others that are located in its blind angle at the back (as is usual in these models). Here, $w_c$ is the fixed weighting factor for cohesion (Equ. 5, Tab. 1). Because of the hard sphere, mentioned above, we ignore cohesion to others within this radius $r_h$ (Equ. 5). The outer borders of flocks of starlings are denser than the interior of the flock (Ballerini et al., 2008b), which is attributed to a kind of defense against predators at the periphery of the group (Hamilton, 1971). Therefore, we make individuals cohere more strongly at the border than in the interior by multiplying by the degree to which an individual is peripheral (Equ. 5, 7). To measure the location of an individual, we use the degree of its 'centrality' in the group, $C_i$ (Equ. 7) (Hemelrijk and Wantia, 2005). It is calculated as the length of the mean vector of the direction towards its neighbouring individuals $N_G$. The 'centrality' $C_i$ has values close to 0 in the interior and values between 0.5-0.75 at the border of a flock. To increase the accuracy of the measurement of the degree of centrality of the individual $C_i$, we base its computation on more than only the number of topological neighbours. We arbitrarily choose to calculate it for the group of 'neighbouring' individuals $N_G$ in a radius of twice the individual's actual perceptual distance (Equ. 7). Note that interactions to cohere with others are not entirely topological, because attraction differs inside and outside the hard sphere.

$$f_{c_i} = C_i \frac{w_c}{|N_i^*|} \sum_{j \in N_i^*} \delta_{ij} d_{ij} \; ; \qquad \delta_{ij} = \begin{cases} 0 & ; x \leq r_h \\ 1 & ; x > r_h \end{cases} \qquad \text{Cohesion (5)}$$

$$N_i^* = \{ \, j \in N_i \, ; \, j \text{ not in the 'blind angle' of } i \, \} \qquad \text{Reduced neighbourhood (6)}$$

$$C_i = \frac{1}{|N_G|} \left\| \sum_{j \in N_G} d_{ij} \right\| ; \qquad N_G = \{ j \in N; d_{ij} \leq 2R_i \} \qquad (7)$$

As regards the alignment behaviour (Equ. 8), individual $i$ perceives a force, $f_{a_i}$, to align with the average forward direction of its $N^*_i$ interaction neighbours (which are the same neighbours as to whom it is attracted).

$$f_{a_i} = w_a \left( \sum_{j \in N_i^*} e_{x_j} - e_{x_i} \right) \Big/ \left\| \sum_{j \in N_i^*} e_{x_j} - e_{x_i} \right\| \qquad \text{Alignment (8)}$$

Here, $e_{x_i}$ and $e_{x_j}$ are the vectors indicating the forward direction of individuals $i$ and $j$ and $w_a$ is the fixed weighting factor for alignment (Tab. 1).

The social force is the sum of these three forces (Equ. 9).

$$F_{Social_i} = f_{s_i} + f_{a_i} + f_{c_i} \qquad \text{Social net force (9)}$$

Furthermore, to limit the three dimensional movement to the flight area above the roosting site (Ballerini et al., 2008a; Carere et al., 2009) we made the individuals perceive an attraction force $f_{Roost}$ to this 'roosting area' (Equ. 10). It consists of a horizontal and vertical attraction to the roost (Equ. 11, 12). The strength of the horizontal attraction, $f_{RoostH}$, depends on the degree to which the individual heads outwards when it is not over the roost. It is strong if an individual moves away from the roosting area, and it is weaker if it is already returning. The strength is calculated using the dot product, i.e. the angle between the forward direction of individual $i$, $e_{x_i}$, and the horizontal outward-pointing normal $n$ of the boundary. The range of the result $[-1..1]$ is transformed to $[0..1]$ by halving the dot product and summing it with a 1/2. The actual direction of the horizontal attraction force to the roost is given by $e_{y_i}$ which is the individual's lateral direction. The sign in Equ. 10 is chosen in such a way that the outward heading is reduced. Vertical attraction, $f_{RoostV}$, is proportional to the vertical distance from the preferred height above the roost (arbitrarily called the zero level). Here $z$ is the vertical unit vector. $w_{roostH}$ and $w_{roostV}$ are fixed weighting factors for staying close to the roost.

$$f_{Roost_i} = f_{RoostH_i} + f_{RoostV_i} \qquad \text{Roost boundary force (10)}$$

$$f_{RoostH_i} = \pm w_{RoostH} \left( \frac{1}{2} + \frac{1}{2}(e_{x_i} \cdot n) \right) \cdot e_{y_i} \qquad \text{Horizontal attraction to roost area (11)}$$

$$f_{RoostV_i} = -w_{RoostV} \, (vertical\ distance) \cdot z\, ; \quad z = (0,0,1)^T \qquad \text{Vertical attraction to roost area (12)}$$

The sum of the social force, the forces that control speed and ranging and a random force (indicating unspecified stochastic influences, Equ. 13) is labelled as the 'steering force' (Equ. 14).

$$f_{\xi_i} = w_\xi \cdot \xi \qquad \text{Random force (13)}$$

$$F_{Steering_i} = F_{Social_i} + f_{\tau_i} + f_{Roost_i} + f_{\zeta_i} \qquad \text{Steering force (14)}$$

Here $\xi$ is a random unit vector from a uniform distribution and $w_\xi$ is a fixed scaling factor.

As regards the flying behaviour, we follow the standard equations of fixed wing aerodynamics, as is common in studies of bird flight (Norberg, 1990). During horizontal flight with constant cruising speed $v_0$ the lift $L_0$ balances the weight $mg$ of the bird and the bird generates thrust $T_0$ that balances the drag $D_0$ (Equ. 15)

$$L_0 = \frac{1}{2}\rho S v_0^2 C_L = mg\,; \qquad D_0 = \frac{1}{2}\rho S v_0^2 C_D = T_0 \qquad \text{Default lift and drag (15)}$$

$$L_i = \frac{v_i^2}{v_0^2} L_0 = \frac{v_i^2}{v_0^2} mg\,; \qquad D_i = \frac{C_D}{C_L} L \qquad \text{Lift and drag (16)}$$

$$F_{Flight_i} = (L_i + D_i + T_0 + mg)\,;\ L_i = L_i \cdot e_{z_i}\,;\ D_i = -D_i \cdot e_{x_i}\,;\ T_0 = T_0 \cdot e_{x_i} \quad \text{Flight force (17)}$$

Here, $g$ is the standard gravity, $g=(0,0,g)$ is the gravity vector, $v$ is the speed, $\rho$ is the air density, S is the wing area, $C_L$ and $C_D$ are the dimensionless lift and drag coefficients; $T_0, L_0$, and $D_0$, respectively represent the default thrust, lift and drag at cruise speed $v_0$.

By combining the first two equations (Equ. 15), the lift force can be calculated as a function of the cruise speed, the actual speed and the weight of the individual. (Thus, we need neither $\rho$ nor $S$ to calculate lift and drag.) Note that we assume that weight, wing area, and the dimensionless lift and drag coefficients are fixed and the same for all individuals. The drag is a function of lift and the two coefficients (Equ. 16). $L_i$ and $D_i$ are the actual lift and drag. The total force of flying, the 'flight force', $F_{Flight}$ is calculated by summing the lift, drag, the default value of thrust and the weight (Equ. 17).

When flying along a curve, birds usually roll their wings into the direction of the turn until they are at a certain angle to the horizontal plane, the so-called banking angle (Videler, 2005). This causes the lift to bank also, so that its horizontal component acts as centripetal force which helps to maintain the curved flight path (Warrick et al., 2002). To represent these banked turns we first calculate the degree to which individuals want to turn, or their lateral acceleration, $a_l$, which is exerted by the steering force (Fig. 3). Following the first law of Newton ($\boldsymbol{F}=m\boldsymbol{a}$), we obtain the lateral acceleration

$$\boldsymbol{a}_{l_i} = \left( \frac{\boldsymbol{F}_{Steering_i} \cdot \boldsymbol{e}_{y_i}}{m} \right) \cdot \boldsymbol{e}_{y_i} \qquad \text{Lateral acceleration (18)}$$

$$\tan(\beta_{in_i}) = w_{\beta_{in}} \|\boldsymbol{a}_{l_i}\| \Delta t \qquad \text{Roll in angle (19)}$$

$$\tan(\beta_{out_i}) = w_{\beta_{out}} \sin(\beta_i) \Delta t \qquad \text{Roll out angle (20)}$$

$$\beta_i(t + \Delta t) = \beta_i(t) + \beta_{in_i} - \beta_{out_i} \qquad \text{Banking angle (21)}$$

where $\beta_i$ is the actual banking angle, $w_{\beta in}$ and $w_{\beta out}$, respectively are the weights for rolling in and out the curve of turning, $\Delta t$ is the update time and $\beta_{in}$ and $\beta_{out}$ are the angles over which an individual intends to move inwards and outwards. The extent of lateral acceleration determines the extent of inwards banking. Banking implies that the

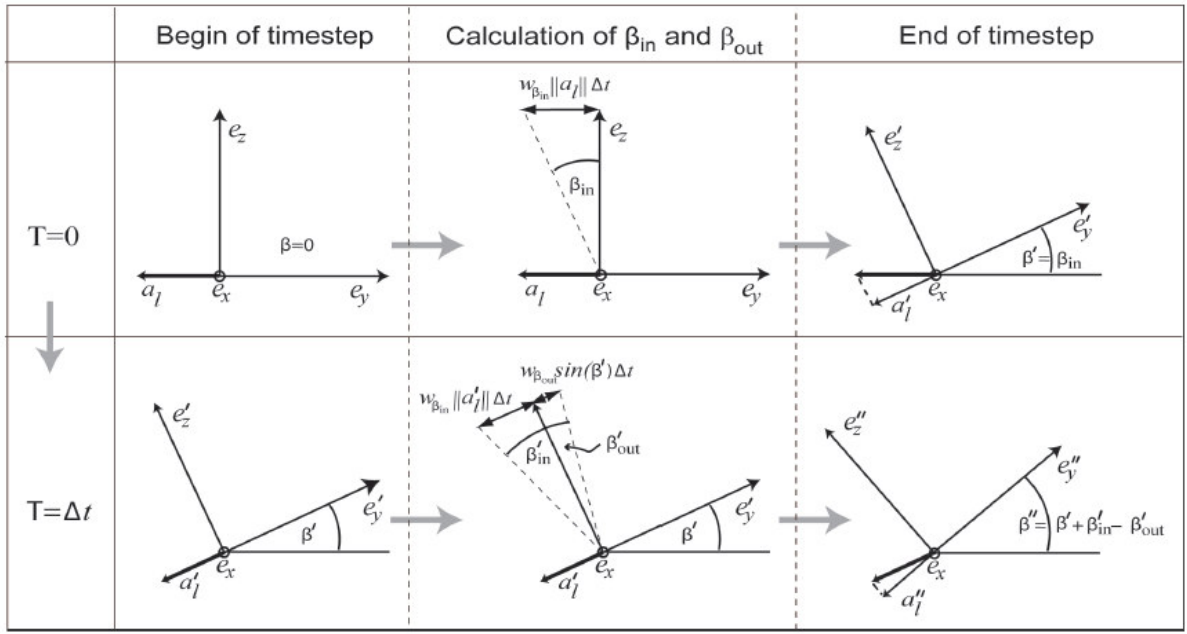

**Figure 3.** Rotation of the individual around its forward direction during banked turns. The forward direction of the individual $e_x$ is pointing towards the reader. Top row, left: at the beginning of the time step (T=0), the individual flies at level flight and it experiences a steering force to turn left with a sideward horizontal component $a_l$. At the end of the time step (top row, right) it banks inwards over an angle $\beta$ in order to take the turn. Bottom row, left: At the next time step, the individual experiences a tendency to both turn inwards $\beta_{in}$ and outwards $\beta_{out}$. Its actual turning angle is given by the sum of its present angle $\beta'$ and both tendencies to turn inwards $\beta'_{in}$ and outwards $\beta'_{out}$. For illustration, we used an imaginary value for $w_{bin} * \Delta t \sim 1$ that differs from the default value. For further description see text.

individual's local frame rolls around the forward axis in the direction of the lateral acceleration, $a_l$. We obtain the intended inward angle as the product of the weight for inward turning, the strength of the lateral acceleration and the update time over which it worked (Equ. 19, Top row in Fig. 3). To make the animal return again to level flight, we also gave individuals a tendency to roll out, which is larger when the actual banking angle is greater (Equ. 20). The actual banking angle is the sum of the current angle and the tendencies to roll-in and to roll-out (bottom row of Fig. 3). The ratio of $w_{\beta in}$ and $w_{\beta out}$ determines the roll rate, or the speed with which it banks and returns to level flight. Note that the acceleration, which was originally strictly lateral, has after the roll

| Parameter | Description | Default value |
|---|---|---|
| $\Delta t$ | Integration time step | 5 ms |
| $\Delta u$ | Reaction time | 50 ms |
| $v_0$ | Cruise speed | 10 m/s |
| $M$ | Mass | 80 g |
| $C_L/C_D$ | Lift-drag coefficient | 3.3 |
| $L_o$ | Default lift | 0.78 N |
| $D_0, T_0$ | Default drag, default thrust | 0.24 N |
| $w_{\beta in}$ | Banking control | 10 |
| $w_{\beta out}$ | Banking control | 1 |
| T | Speed control | 1 s |
| $R_{max}$ | Max. perception radius | 100 m |
| $n_c$ | Topological range | 6.5 |
| S | Interpolation factor | 0.1 $\Delta u$ |
| $r_h$ | Radius of max. separation ("hard sphere") | 0.2 m |
| $r_{sep}$ | Separation radius | 4 m |
| $\Sigma$ | Parameter of the Gaussian g(x) | 1.37 m |
| $w_s$ | Weighting factor separation force | 1 N |
|  | Rear "blind angle" cohesion & alignment | 2*45° |
| $w_a$ | Weighting factor alignment force | 0.5 N |
| $w_c$ | Weighting factor cohesion force | 1 N |
| $C_c$ | Critical centrality below which an individual is assumed to be in the interior of a flock. | 0.35 |
| $w_\xi$ | Weighting factor random force | 0.01 N |
| $R_{Roost}$ | Boundary radius | 150 m |
| $w_{RoostH}$ | Weighting factor horizontal boundary force | 0.01 N/m |
| $w_{RoostV}$ | Weighting factor vertical boundary force | 0.2 N |

**Table 1**. Default parameter values.
Note that $T_0$ is calculated by equation 15.

received, parallel to the lift, a component of magnitude $\|a_l\| \sin(\beta_{in})$, shown as dotted

lines in the last column of Fig. 3 . This converts lateral acceleration into additional lift which compensates for the loss of effective lift due to banking as has been observed in real birds (Müller and Lentink, 2004; Pomeroy and Heppner, 1992).

After summing the forces of steering and flying, we apply Euler integration to

calculate the position and velocity at the end of each time-step $\Delta t$:

$$v_i(t + \Delta t) = v_i(t) + \frac{1}{m}\left(F_{Steering_i}(t) + F_{Flight_i}(t)\right)\Delta t \quad (22)$$

$$r_i(t + \Delta t) = r_i(t) + v_i(t + \Delta t) \cdot \Delta t \quad (23)$$

Here, $v_i$ is the speed of individual $i$, $m$ its mass, $r_i$ its location, and $\Delta t$ is the update time.

**Parameterization & Experiments**

As far as data are available, we parameterise the model to realistic values (Tab. 1).

We base our values for body mass, lift/drag ratio, $C_L/C_D$, and cruise speed on data specific for starlings from studies by Ward *et al.* (Ward et al., 2004) and for reaction time on those for visual stimuli from experiments by Pomeroy and Heppner (Pomeroy and Heppner, 1977). We tune the interpolation factor $s$ (Equ. 2) to obtain the topological range $n(t)$ of 6.5 ±1 neighbours as observed in starlings (Ballerini et al., 2008a).

We scale the steering force (i.e. by tuning the weights for repulsion, alignment and attraction) to have average values around 1Newton, which is the order of magnitude of the physical forces of flight (see $L_0$ and $D_0$ in Tab. 1). We set the hard sphere or minimal distance, $r_h$, to starlings' average wingspan (Ballerini et al., 2008b; Möller, 2005; Ward et al., 2004). The weight of the separation force, $w_s$, is tuned in such a way that collisions never occurred (Potts, 1984) . We adjust the average nearest neighbour distance (using radius of separation $r_{sep}$) to that of the starlings of Rome (Tab. 2). Note that the average interaction radius and the nearest neighbour distance increase linearly with the radius of separation (Fig. 4).

We adjust the horizontal attraction to the roost area (tuning $w_{roostH}$) in order to limit their movements to a horizontal area that resembled the roost in Rome with radius, $R_{Roost}$ (Tab. 1) and tune the vertical attraction to prohibit them from flying off vertically and to produce the 'flat' flock shape observed in real flocks (calibrating $w_{roostV}$) (Ballerini et al., 2008b). In order to match empirical data regarding the angle under which individuals keep their nearest neighbour, the so-called bearing angle, individuals also avoid others in their blind angle, thus they avoid each other all-around (Fig. 2). This is reasonable, since starlings frequently scan by head turning while flying (personal observations) and it is very dangerous for them to collide with each other. We retain the blind angle for the forces of alignment and cohesion, because it is difficult to react to others at its back. Therefore, we suppose that the bird reacts to others at its back only for behavioural acts for which this is really necessary. We adjust the flying behaviour by observing the behaviour of a single individual in flocks of several sizes. We tune the relaxation parameter of the speed $\tau$ in combination with the vertical attraction to the roost $w_{RoostV}$ so that flocks stay at level flight when they are not disturbed (by having to turn or merge). The reaction time of individuals is represented in the update time of the 'steering' force ($\Delta u$ seconds). We make this interval longer than the integration time step of location and speed (of $\Delta t$ seconds, see Equ. 22,23), which causes minor errors in their navigation. To further reflect the fact that decision making in animals is subject to stochastic effects we also add a random force, which we scale to 10% of the steering force.

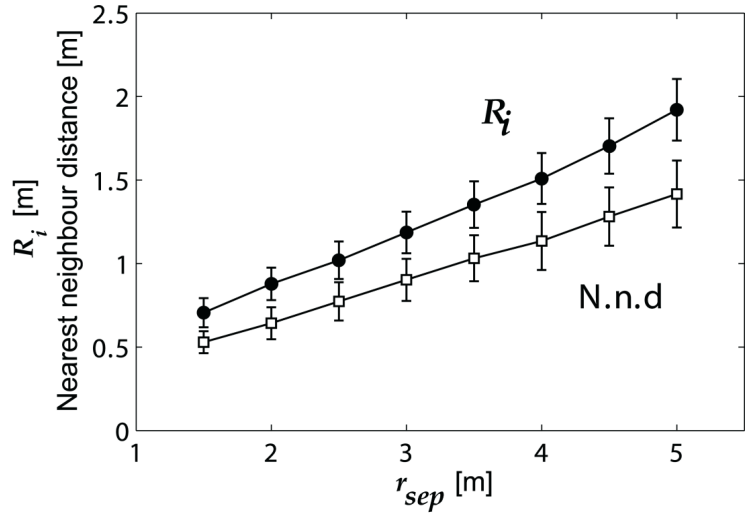

**Figure 4.** The relation between the size of the separation radius $r_{sep}$ and the nearest neigbour distance (NND, squares) and the actual interaction radius (Ri, circles) indicated in average and standard deviation when flying over the roost. For each value of the separation radius data were collected at 10 second intervals during a period of 30 min. N=2000, apart from $r_{sep}$ default parameters were used, see Tab 1.

We have made side-by-side comparisons of the visualisation of the model and video recordings (made by Claudio Carere). These video recordings were made at the roost close to the central railway station 'Termini' in Rome, Italy. We choose this site because predators, which tend to induce more complicated collective responses, were more rarely observed here than at the other main roost in Rome (Carere et al., 2009). The flocks were filmed between 19[th] January 2006 and 17[th] March 2006 and between 12[th] December 2006 and 2[nd] March 2007 with a HD video camera (JY-HD10, JVC). We have used video recordings to tune our model qualitatively in two ways, 1) to adjust the parameters controlling banking, $w_{\beta in}$ and $w_{\beta out}$, to match the roll rates and bank angles as observed in the videos of real flocks  2) to fine-tune the weight of the alignment force, $w_a$ , so as to reproduce the high degree of polarization observed in the videos of starlings.

At the beginning of each simulation individuals have been released at random positions and with random orientations inside a cylinder with radius, $R_{Roost}$, which represents the approximate size of the roost at termini in Rome (Carere et al., 2009).

**Measurements**

Below we describe the measurements of the model. These we made in a similar way to those in empirical data that were collected in Rome (Ballerini et al., 2008a; Ballerini et al., 2008b).

**Distances to nearest neighbours**

Nearest neighbour distances are measured 10 times per second and averaged per flock (Tab. 2). To avoid border effects only individuals in the interior of the flock are considered (with centrality < 0.35).

**Angular distribution of nearest neighbours**

The angular distribution of nearest neighbours of all individuals is measured as the distribution of cosines of the angle $\theta$ between the forward direction of an individual and the vector to its nearest neighbour. A cosine $\theta$ of +1 indicates that the nearest neighbour is ahead, a cosine of 0 indicates the nearest neighbour is at an angle of 90 degrees on either side and a cosine of -1 indicates that it is at the back of the individual in question. Nearest neighbour directions have been measured 10 times per second and averaged per flock (Tab. 1). To avoid the influence of border effects only individuals in the interior of the flock are considered (i.e. individual with centrality $C_i$ < 0.35).

**Volume**

The volume of a flock is measured by mapping the position of the individuals on a cubic lattice and counting the occupied lattice cells, so-called voxelisation. The cell size has been adapted to the average distance to the nearest neighbours in the flock.

**Balance shift**

Balance shift (Ballerini et al., 2008b) is computed as the difference between centre of mass and geometrical centre divided by the length of the axis along the direction of motion. Positive values indicate higher frontal density. The balance shift is recorded every 0.1s over 2 minutes (1200 samples).

**Border and interior of flock**

Border and interior of flock are defined as the volume where centrality $C_i > 0.35$ and $C_i <= 0.35$ respectively. Data have been collected at 30 second intervals over 30 min.

**Aspect ratio**

The aspect ratio is measured by means of the approximated minimum-volume bounding box of the flock, calculated by means of a principal component analysis (PCA) of the coordinates of the flock members (Barequet and Har-Peled, 2001)(movie S3). The eigenvectors that are associated with the largest/medium/smallest eigenvalue of the covariance matrix provide a coordinate system oriented along the longest/medium/shortest axis of the flock, respectively, $I_3, I_2, I_1$.

**Orientation parameters**

Following the analysis of the empirical data, orientation parameters are given by three angles (Ballerini et al., 2008b): the angle between the shortest dimension of the flock and gravity, the shortest dimension and the direction of speed, and the angle between the velocity of the flock and gravity. These are calculated by the absolute value of the cosine of the angle, respectively between the shortest dimension of the flock vector $I_1$ and gravity $G$, $|I_1 \cdot G|$, vector $I_1$ and normalised velocity $V$ $|V \cdot I_1|$ and between the vectors of normalised velocity $V$ and gravity $G$, . $|V \cdot G|$

**Flock banking**

Flock banking is defined as the rotation angle of the flock around its direction of movement against the horizontal plane. Given that $I_1$ is the 'up'-direction of the bounding box and the velocity of the flock's centre of mass is $V$, the banking of the flock is the angle between the horizontal plane and the vector orthogonal to the plane of $I_1$ and $V$ (thus $I_1$x$V$, where 'x' indicates the cross-product). This is calculated every 0.1 second for 8 minutes.

**Results**

In the model flocks emerge. First, we show the resemblance of the full model to empirical data. For this, we compare modeling data to qualitative (Carere et al., 2009) and quantitative data of starlings in Rome. Quantitative data have been obtained by Ballerini and co-authors using stereophotography (Ballerini et al., 2008a; Ballerini et al., 2008b). Second, we show that the starling like patterns in the model depend on 1) banking while turning, 2) group size, 3) the maximal repulsion within the diameter of the wing span, 4) extra attraction at the border of the flock and 5) all round repulsion.

**Comparison of model and qualitative empirical data**

The results of our model resemble the empirical data qualitatively concerning 1) the shapes of the flocks (Fig. 5A), 2) the way they move and split into sub-flocks (Fig. 5B), their changes in shape and density over time (Movies 1,2 of the model vs. Movie S8), the trajectories of individuals during a turn (Fig. 6A) and 5) the way complete flocks tilted sideward while they are turning (Fig. 6B). During turns of a small flock in the model, the trajectories of all individuals are similar in length and in radius (Fig. 6A, movie S5). This means that the individuals after the turn end up in another location inside the flock. For instance, during the right turn in Fig. 6A the individual indicated by a square is initially located at the left side of the flock (at T=0s) but after the turn (at T=3s) it is located at the right side. Note that in this case the shape of the flock is preserved. This can be seen from the triangle connecting the three marked individuals. Its shape remains approximately intact at all three points in time, but its orientation has changed relative to the movement direction. This can be seen from the fact that the point of the triangle initially points backwards (T=0s), but after the turn (3, 6 seconds) it points forward. Sometimes during turns of flocks in the model the complete flock banked in the direction of individual banking or tilted in other directions. Although it happened most often in the direction of individual banking (thus in the direction of the turn) it occurred also in other directions as can be observed from the large variance of data of individual banking and flock banking of flock M32-06 (Fig. 6B). In this flock, its degree of tilting was positively correlated with the average degree of banking by the flock members (Pearson correlation, N=4800, r=0.71, P<0.001). This happens sometimes but only for spontaneous turns of small flocks, not for turns that are induced by attraction to the roost when crossing its border.

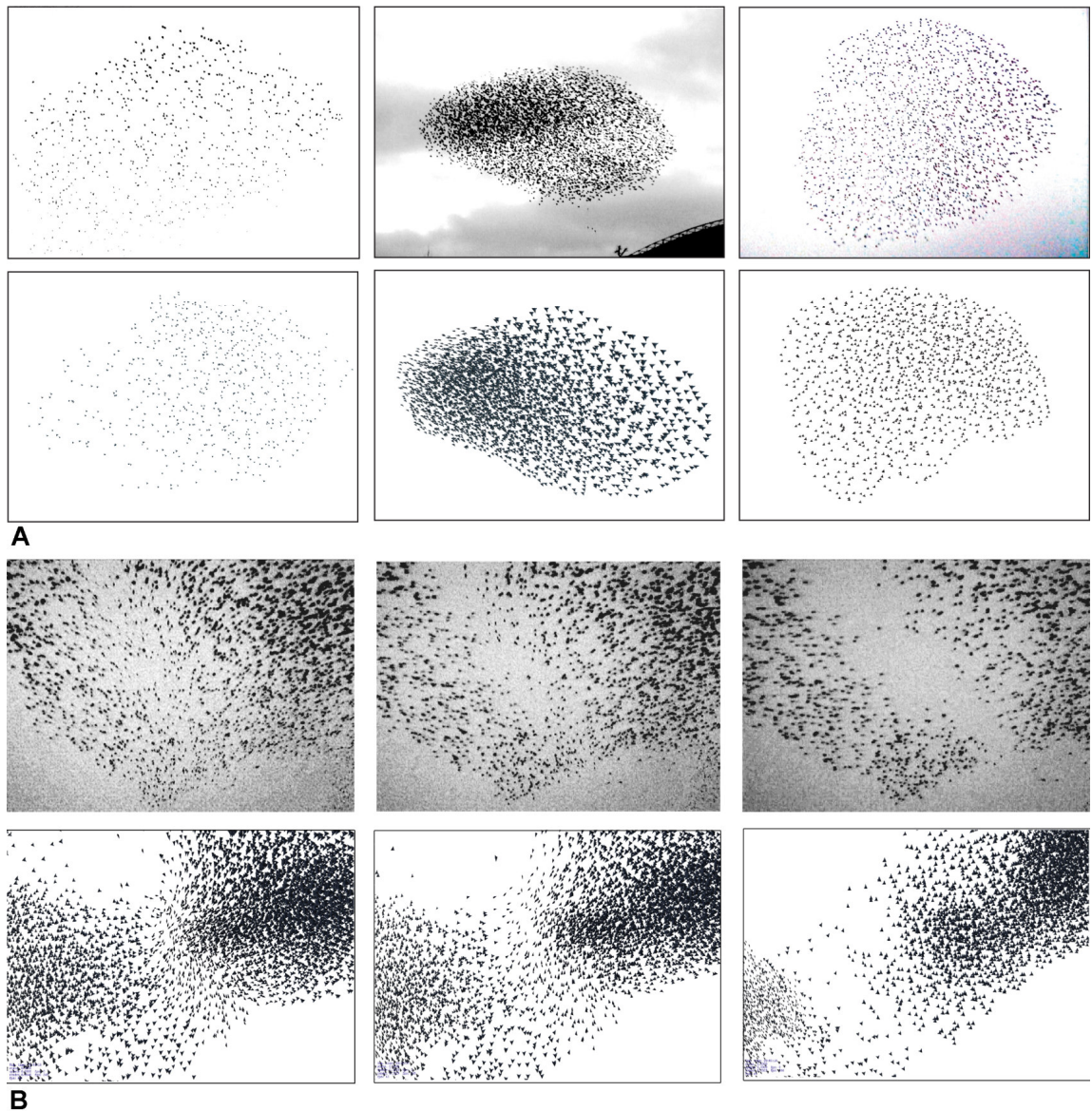

**Figure 5.** Images of A) flocks and B) of the process of splitting a flock. A) Images of flocks of different sizes and densities from video recordings in Rome (Top row) and in the model (Bottom row). Estimated flock size (from left to right): N=700, N=2500-3000, N=2000. The outer flocks move straight forward while the one in the middle is turning. To match the sparse flock at the left, the separation radius in the model was set to $r_{sep}$=5. Otherwise, the parameters are kept at their default value. B) A close-up of the process of a flock split in video recordings in Rome (top row) and in the model (bottom row). The time interval between each picture (in each row) is 1 s.

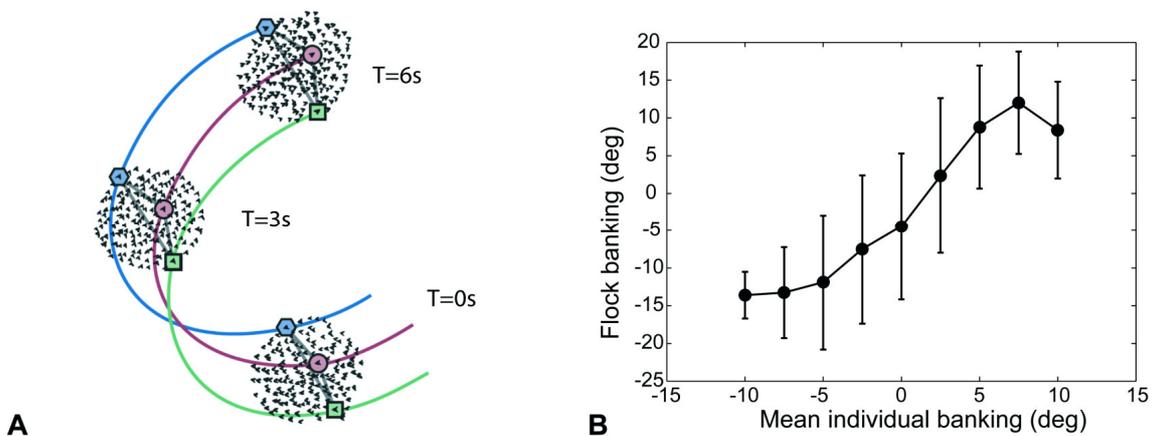

**Figure 6.** Properties of flocks, turning behaviour and internal structure: A) Right-turn of a small flock in the model (N=50, default parameters). The trajectories of three individuals are marked at three points in time. B) Flock banking versus average individual banking in the model (average and standard deviation for flock M3206).

**Comparison of model to quantitative empirical data**

Global quantitative properties were measured in ten flock events in Rome which differed in their size, density and speed (Ballerini et al., 2008b). These measurements included the balance shift, i.e. distance between the geometrical centre and the centre of mass. This was used to determine whether or not there was an imbalance in density between the front and the back. When its value was small the density was similar at the front and back; The shape of the flock was measured in terms of its aspect ratios; its direction of movement relative to the ground was given by the orientation parameters relative to gravity and to the ground; The distribution of distances and angles towards the nearest neighbours was recorded. To compare modelled flocks to empirical ones we first tuned in the model flock size ($N$), speed (by tuning cruise speed $v_0$) and average distance to the nearest neighbour (by tuning the separation radius $r_{sep}$) to those of each empirical flock event. Next we studied whether flocks emerged in the model that resembled the empirical ones in eleven other traits studied in Rome (Tab. 2, Fig. 7ABC). For each ten empirical flock events we found flocks in the model that were similar for many traits.

As to the other traits measured of the ten flock events, the balance shift in the model did not differ from the empirical data (Wilcoxon matched-pairs signed-ranks test, N=10, sr=24.5, NS). It is low, which indicates that the density is similar at front and back half of the flock (Tab. 2).

The shape of a flock is measured through the ratios of the two longer dimensions of the flock, $I_2$ and $I_3$ to the shorter one, thus, $I_2/I_1$ and $I_3/I_1$. These ratios did not differ from empirical values (Wilcoxon matched-pairs signed-ranks, for $I_2/I_1$ N=10, sr=20, NS; for $I_3/I_1$, N=10, sr=26, NS); they were similar between flocks (Tab. 2). Like in empirical data shape as measured by the longest divided by the shortest dimension $I_3/I_1$ did not depend on flock size or volume (Kendal tau between $I_3/I_1$ and flock size, $\tau$ =0.33, NS; $I_3/I_1$ and flock volume, $\tau$ =-0.07 NS), but flock shape measured by the other aspect ratio $I_2/I_1$ did depend on flock size and volume (Kendal tau between $I_2/I_1$ and flock size, N=10, $\tau$ =0.51 P=0.05, $I_2/I_1$ and flock volume, $\tau$ =0.51, P=0.05). Here, it appears that the quotient of the two shorter dimensions increases with flock size. In other words larger flocks become somewhat more asymmetric, smaller ones are closer to being spherical (see also figure 9).

| Flock/ Flocking event | Number of Birds | rSep (m) | Cruise Speed (m/s) | Velocity (m/s) | NND (m) | Balance shift | Volume ($m^{-3}$) | Thickness $I_1$ (m) | Aspect ratios | | Orientation parameters | | |
|---|---|---|---|---|---|---|---|---|---|---|---|---|---|
| | | | | | | | | | $I_2/I_1$ | $I_3/I_1$ | $|I_1 \cdot G|$ | $|V \cdot G|$ | $|V \cdot I_1|$ |
| M32-06 (E32-06) | 781 | 1.6 | 10.0 | 9.8 (9.8) | 0.69 (0.68) | 0.04 (0.08) | 532 (930) | 4.27 (5.33) | 2.02 (2.97) | 4.60 (4.02) | 0.85 (0.89) | 0.01 (0.06) | 0.49 (0.20) |
| M28-10 (E28-10) | 1246 | 1.75 | 11.0 | 10.8 (10.9) | 0.70 (0.73) | 0.1 (-0.06) | 1500 (1840) | 4.35 (5.21) | 2.94 (3.44) | 6.67 (6.93) | 0.97 (0.80) | 0.04 (0.09) | 0.20 (0.41) |
| M25-11 (E25-11) | 1168 | 2.0 | 9.0 | 8.8 (8.8) | 0.78 (0.79) | 0.1 (-0.1) | 1781 (2340) | 5.87 (8.31) | 2.15 (1.90) | 5.40 (5.46) | 0.99 (0.92) | 0.03 (0.12) | 0.08 (0.14) |
| M25-10 (E25-10) | 834 | 2.4 | 12.0 | 12.1 (12.0) | 0.89 (0.87) | -0.06 (0) | 1609 (2057) | 5.41 (6.73) | 2.30 (2.65) | 4.80 (4.98) | 0.96 (0.99) | 0.01 (0.18) | 0.16 (0.18) |
| M21-06 (E21-06) | 617 | 3.0 | 12.0 | 11.9 (11.6) | 1.10 (1.00) | 0.04 (0) | 2375 (2407) | 6.42 (7.23) | 2.18 (2.56) | 4.41 (4.53) | 0.99 (0.96) | 0.04 (0.09) | 0.01 (0.11) |
| M29-03 (E29-03) | 448 | 3.0 | 10.1 | 10.2 (10.1) | 1.00 (1.09) | 0.04 (0) | 1218 (2552) | 4.66 (6.21) | 1.97 (3.58) | 4.31 (5.96) | 0.98 (0.97) | 0.02 (0.27) | 0.08 (0.06) |
| M25-08 (E25-08) | 1360 | 4.5 | 11.9 | 11.9 (11.9) | 1.26 (1.25) | 0.02 (0.16) | 5171 (12646) | 6.00 (11.92) | 3.95 (3.32) | 5.65 (5.12) | 0.98 (0.95) | 0.01 (0.14) | 0.12 (0.12) |
| M17-06 (E17-06) | 534 | 4.0 | 9.5 | 9.1 (9.2) | 1.28 (1.30) | 0.02 (0.5) | 3483 (5465) | 6.46 (9.12) | 2.75 (2.76) | 6.10 (6.94) | 0.98 (0.91) | 0.03 (0.09) | 0.08 (0.32) |
| M16-05 (E16-05) | 2631 | 4.0 | 15.0 | 15.0 (15.0) | 1.33 (1.31) | 0.04 (0) | 17388 (28128) | 13.04 (17.14) | 2.36 (2.46) | 4.13 (4.07) | 0.98 (0.90) | 0.01 (0.19) | 0.20 (0.25) |
| M31-01 (E31-01) | 1856 | 5.4 | 6.9 | 7.0 (6.9) | 1.54 (1.51) | 0.05 (0.17) | 11547 (33487) | 6.57 (19.00) | 6.09 (2.44) | 8.646 (4.07) | 0.99 (0.95) | 0.05 (0.09) | 0.03 (0.13) |

**Table 2.** Quantitative properties of flocks in the model and (between parentheses) in empirical data in Rome (Ballerini et al., 2008b). Empirical data of flocks are labelled with an 'E', e.g. E32-06, flocks of our model start with 'M'. The numbers correspond to the empirical observations in Tab. 1 (Ballerini et al., 2008b). $r_{sep}$: radius of separation, Velocity: Speed of the centre of mass. NND: average nearest neighbour distance, Balance shift: Positive values indicate higher frontal density, $I_1$, $I_2$, $I_3$: shortest /medium/ longest axis of the flock. G: unit vector parallel to gravity, V: unit vector of the normalised velocity. Left to the fat line parameters have been tuned to empirical data.

Like in the empirical data, the orientation of the flocks in space was mostly horizontal and parallel to the ground. This was measured by studying to what extent the shortest dimension of the flock $I_1$ was parallel to the gravity G and thus how close the absolute value of their dot product $|I_1 \cdot G|$ was to 1. Although this dot product differed significantly between the flocks in the model and in the empirical data (Wilcoxon matched-pairs signed-ranks test, N=10, sr=6, P=0.03 two-tailed), the average values

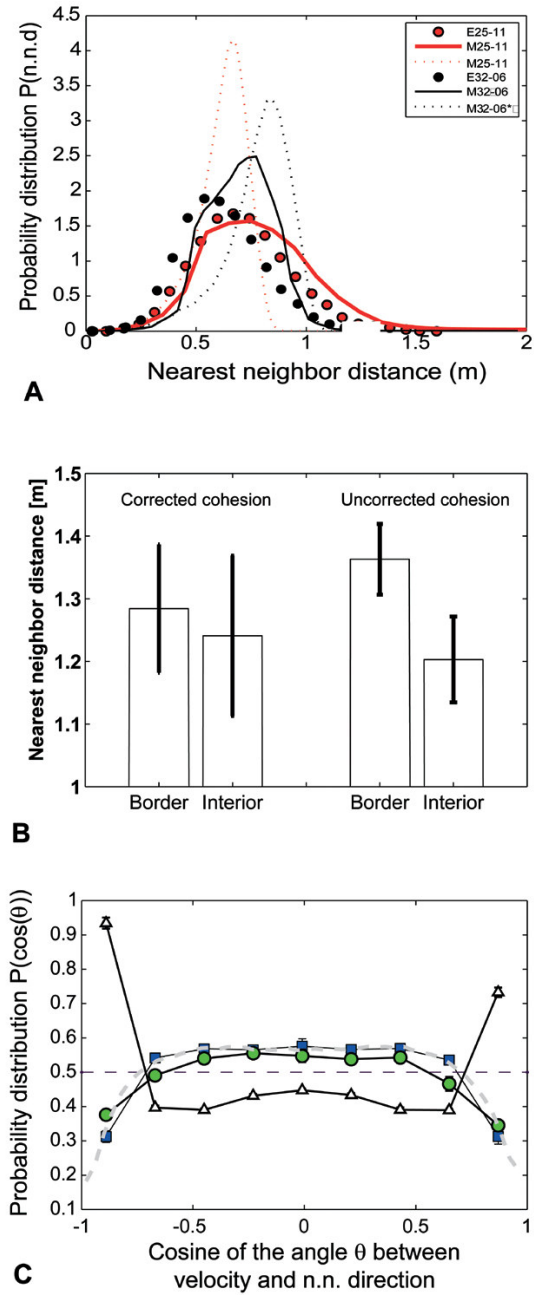

**Figure 7.**

**A)** Distribution of nearest neighbour distance of two flocks that differ in their average density (Tab. 1). In red: closed circles: empirical data from flock E25-11, solid line: corresponding flock in the model M25-11 with hard sphere, dotted line: the same M25-11 without hard sphere and with a separation radius $r_{sep}$=3.5 (instead of 2.0). In black: closed circles: empirical data from flock E32-06, solid black line: corresponding flock M32-06 with hard sphere, dotted black line: the same M32-06 without hard sphere and with a separation radius $r_{sep}$ =2.6 (instead of 1.6).
**B)** Average nearest neighbour distance at the interior of the flock and at its border (average and standard deviation) as measured with extra attraction at the border ('corrected version' at the left) and without this correction (at the right). Flock size was 2000; for the uncorrected case, the factor $C_i$ in Equ. 5 (calculation of the separation force) was kept constant with a value of 0.35 in order to obtain identical overall nearest neighbour distances; apart from this, default parameter were used (Tab. 1).
**C)** Distribution of directions of nearest neighbour (i.e. 'bearing angles'). Green circles: average angular distribution of the corresponding empirical data E21-06, E32-06, E25-10 and E25-11 after (Ballerini et al., 2008b). Blue rectangles: average angular distribution of the modelled flocks M21-06, M32-06, M25-10 and M25-11. White triangles: average angular distribution of modelled flocks of individuals with a blind angle at the back. Grey curve: averaged nearest neighbour distribution of 100 randomly chosen flocks in the model. Dashed line: expected value for a random isotropic system $P_{random}(\cos(\theta)) = ½$. Error bars: standard error.

were similar $|I_I·G|$ =0.97±0.04 in the model and $|I_I·G|$ =0.92±0.06 in empirical data. In the model flocks moved mostly at a constant height above the ground, in other words the velocity of the centre of mass of the flock and the direction of gravity were approximately orthogonal (and thus, the dot product of the normalised velocity and gravity, |**V·G**| was on average low**)**. While this resembles empirical data, the modelled flocks changed their altitude above the ground less (Wilcoxon matched-pairs signed-ranks test, N=10, sr=0, P<0.01 two-tailed) as indicated by the fact that our average |**V·G**| = 0.03 ± 0.04 is closer to zero than is the average empirical value |**V·G**|=0.13 ±0.06 (Ballerini et al., 2008b). The degree to which the flock was oriented parallel to the floor or tilted in any direction (measured by |**V·I**$_1$|, thus the angle between the velocity and the direction of the shortest dimension, $I_1$ ), did not differ significantly from the empirical data (Wilcoxon matched-pairs signed-ranks test, N=10, sr=11, NS), on average in the model it was |**V·I**$_1$| = 0.15 ± 0.14 and in the empirical data |**V·I**$_1$| =0.19 ±0.11. In sum, we conclude that although somewhat stronger than in the empirical data (Ballerini et al., 2008b), our modeled flocks flew on average horizontally and parallel to the ground.

The thickness of the flock is measured along the axis that is on average parallel to gravity **G**. As in the real data thickness in the model is correlated with the volume of the flock (Kendal tau between $I_1$ and volume, N=10, $\tau$ =0.78, P<0.005) and with the number of individuals (Kendal tau between $I_1$ and flock size, N=10, $\tau$ =0.33, NS ) but its absolute values are significantly smaller (Wilcoxon matched-pairs signed-ranks test, N=10, sr=0, P< 0.01 two-tailed). Also the volumes of the modelled flocks are on average significantly smaller than those of the empirical flock events (Wilcoxon matched-pairs signed-ranks test, N=10, sr=0, P<0.01 two-tailed).

The internal structure of flocks is described in the empirical data by the distribution of the distances to the nearest neighbours (for two flocks) and the angles to the nearest neighbours (for four flocks). We compare the modelled flocks to their corresponding empirical flock events. Of the empirical flock events E25-11 and E32-06 (Ballerini et al., 2008b) the frequency distribution of the distances to the nearest neighbours appears to change greatly at around the hard sphere (distance of a wingspan at 0.4 m). Further, the distribution is skewed to the right as measured by the second derivative of the shape of the curve (skewness for E25-11 and E32-06 are respectively, +0.36 and +0.39). In the model of the corresponding flocks, the skew is to the right also (skewness for M25-11 and M32-06 are respectively, +0.37 and +0.54, Fig. 7A).

Like in the empirical data (Tab. 2) also in our model density appears to be independent of group size in the model (Pearson correlation, N=140, r=0.32, P=0.14) if we measure for the default radius of separation $r_{sep}$ (thus without tuning the nearest neighbour distance to empirical data) the nearest neighbour distance in different group sizes for an even larger range of group sizes than the empirical data (namely for a group size of 100, 200, 300, 400, 500, 600, 700, 800, 900, 1000, 2000, 3000, 4000, 5000 individuals) and do so for 10 replicas.

In contrast to empirical findings where the border is denser than the interior of the flock (Ballerini et al., 2008b), in our model, the density was slightly lower at the border than in the centre of the flock (average NND at the border 1.28, in the centre 1.24, Wilcoxon matched pairs signed ranks test, N=60, t=3183, P < 0.05, Fig. 7B).

For empirical data of four flocks, 32-06, 17-06, 25-10 and 25-11, the distribution of the angle between the heading and the nearest neighbour (i.e. 'bearing angle') shows that closest neighbours are lacking along the movement direction (Ballerini et al., 2008b). In our modelling study of the same four flocks we confirm this pattern, it does not differ significantly from the empirical pattern: Its distribution is correlated with the empirical data (Kendall tau, N=9, $\tau$=0.72, P<0-01) and its values do not differ significantly from empirical data (Wilcoxon matched pairs signed ranks test, N=9, sr=14, NS), i.e. there are hardly any nearest neighbours behind (at 180°, cosine -1) and ahead (0°, thus cosine 1) (compare model data in blue squares to empirical data in green circles, Fig. 7C).

The dynamics of these patterns in the model can only be compared qualitatively to data of real starlings, since quantitative data of time-series are still lacking; In the model the degree of banking by individuals is correlated with the curvature of their paths (Fig. 8A); the turn starts by the banking of the individuals (Fig. 8A I-II), leading to loss of effective lift (Fig. 8B I-II) and altitude and thus, increase of vertical speed (Fig. 8B I-II). After passing the apex of the turn, banking is reduced (Fig. 8A II-III) and effective lift is gained (Fig. 8b II-III), while there is still a loss of altitude, by which horizontal speed increases (Fig. 8C II-III). Subsequently, individuals slowly return to level flight (Fig. 8A III-IV), to their original altitude and horizontal speed (Fig. 8C III-IV). While turning, the flock is compressed (Fig. 8D), and although its thickness remains relatively undisturbed, its relative proportions, i.e. its aspect ratio, changes (Fig. 8E). Thus, the variability of flock shape arises in the model mostly during turning and due to this variability the shape of the flock in the direction of movement was often not oblong (movie S3).

**Causes of starling-like patterns**

The banking during turning and the attraction to the roost contribute remarkably to the variability and complexity of the movements of a flock, because if we switch them off variation in effective lift, altitude, horizontal speed, volume and aspect ratios decline (Fig. 10). Further, the flock becomes oblong, and particularly so if individuals do not bank (compare *I3/I1* between Fig. 10E and K). Without banking, the flock moves around the border of the roost in circles and it lengthens until it covers the complete circumference of the roost. Without attraction to the roost, the flock moves straight ahead and is much thicker due to the absence of vertical attraction to a preferred level above the roost (the zero-level). This becomes clear by comparing its thickness without such attraction (Fig. 10K) and with vertical attraction (Fig. 10E and Fig. 8E).

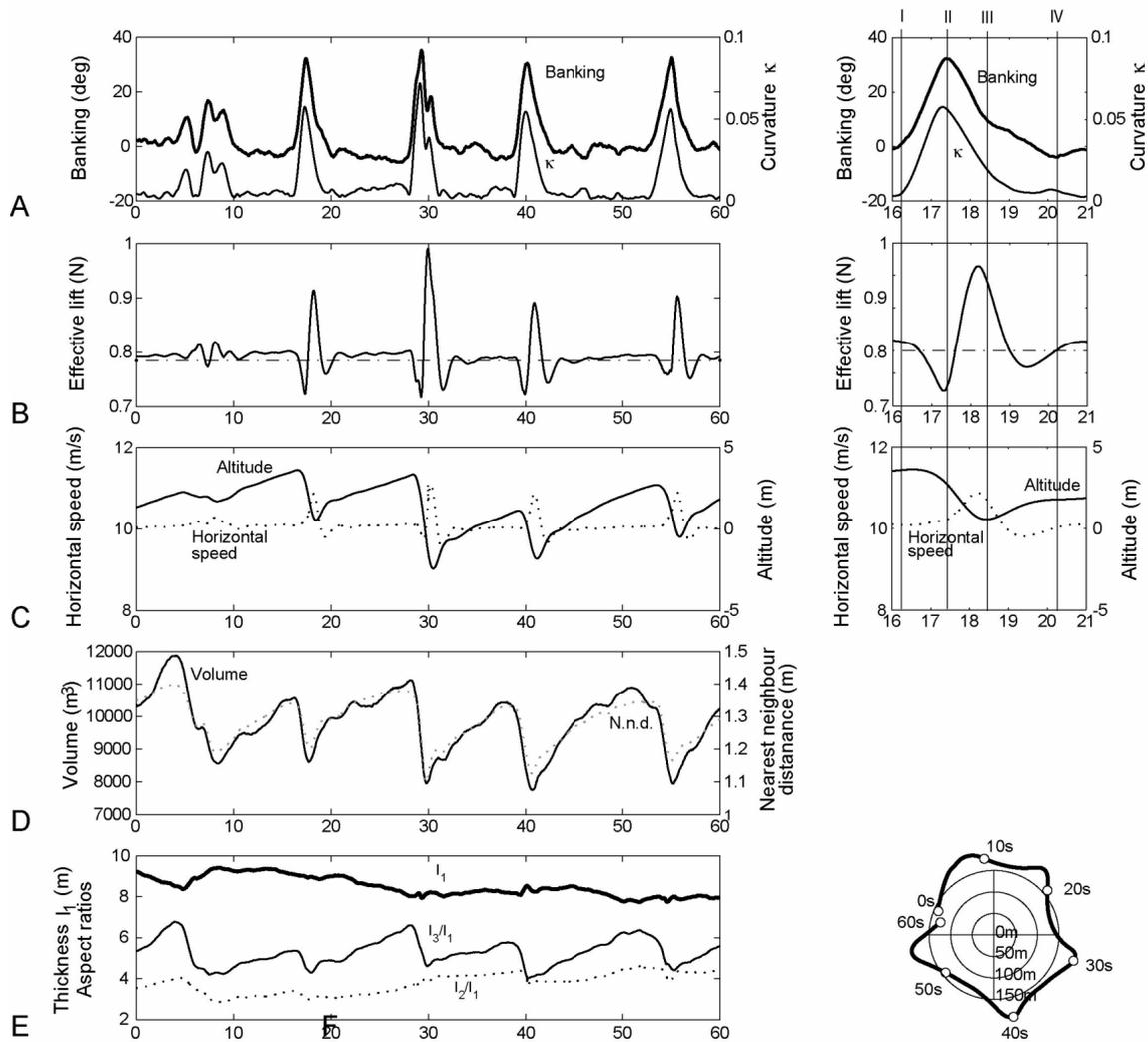

**Figure 8.** Time series of flight behaviour and flock properties in the model. X-axis in seconds. A) average bank angle and path curvature, $\kappa$ ; B) Effective lift, dashed line equals lift at cruise speed and at level flight; C) horizontal speed and altitude (zero is the preferred level above the roost); At the right of A), B), and C) a close-up is shown, in which I, II, III, and IV mark transitions between phases described in the text; D) Volume and nearest neighbour distance (NND); E) Aspect ratio measured as $I_3:I_1$ (largest : shortest extent), $I_2:I_1$ (middle : shortest extent) and thickness ($I_1$). F) trajectories (spatial top view). Numbers indicate points in time. Flock size N=2000, default parameters.

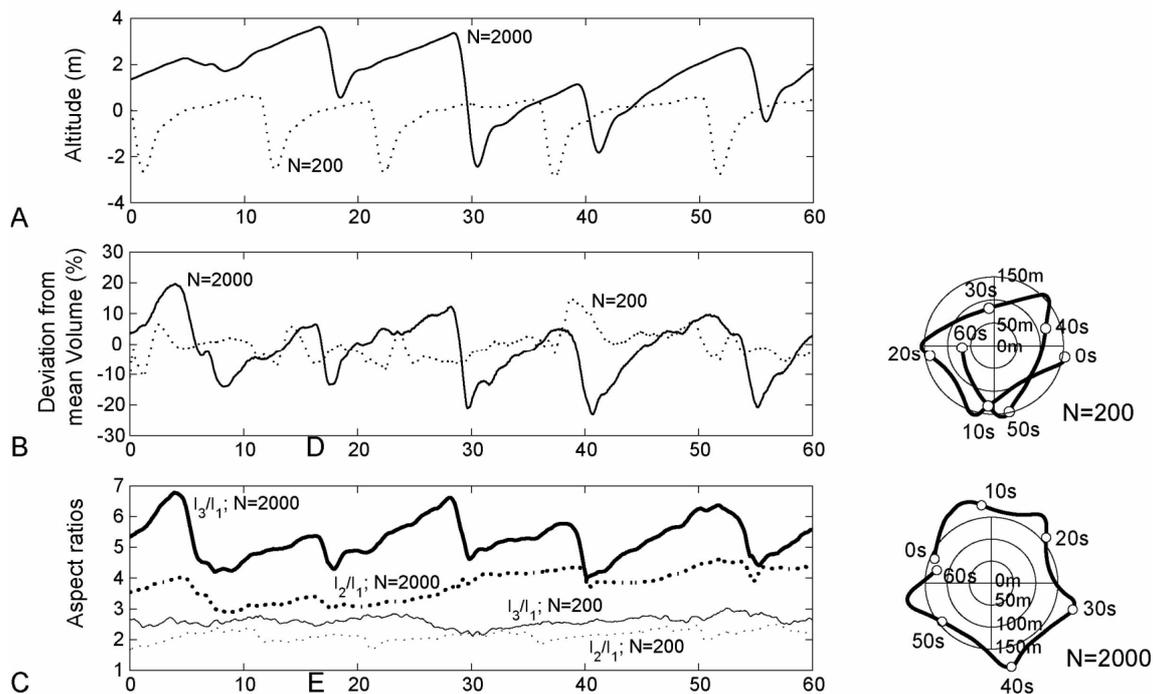

**Figure 9.** Comparison between a small flock of 200 individuals and a large flock of 2000 individuals regarding the time series of flock properties. X-axis in seconds. A) vertical distance above and below the preferred height above the roost (indicated as zero), altitude, B) the percentage with which the flock deviates from its average volume, C) aspect ratios measured as $I_3:I_1$ (largest : shortest extent), $I_2:I_1$ (middle : shortest extent).

Compared to the larger flocks (of above 500 individuals in Tab. 2), small flock size (200 individuals) reduces the complexity of the displays in terms of: 1) the variation in vertical distance (i.e. altitude), 2) the proportional variation in volume and 3) the variation in the aspect ratios (Fig. 9). Further, the shape of the small flock is more spherical as reflected in its lower aspect ratios ).

The typical right skew of the distances to the nearest neighbour (measured above), which resembles empirical data, is due to the hard sphere, because it disappears if we take the hard sphere out. Without this hard sphere the distribution of nearest neighbour distances produces a skew in the opposite direction, to the left, for M25-11 and M32-06, skewness is respectively -0.92 and -1.02 (see dotted red line for M25and dotted black line for M32, Fig. 7A).

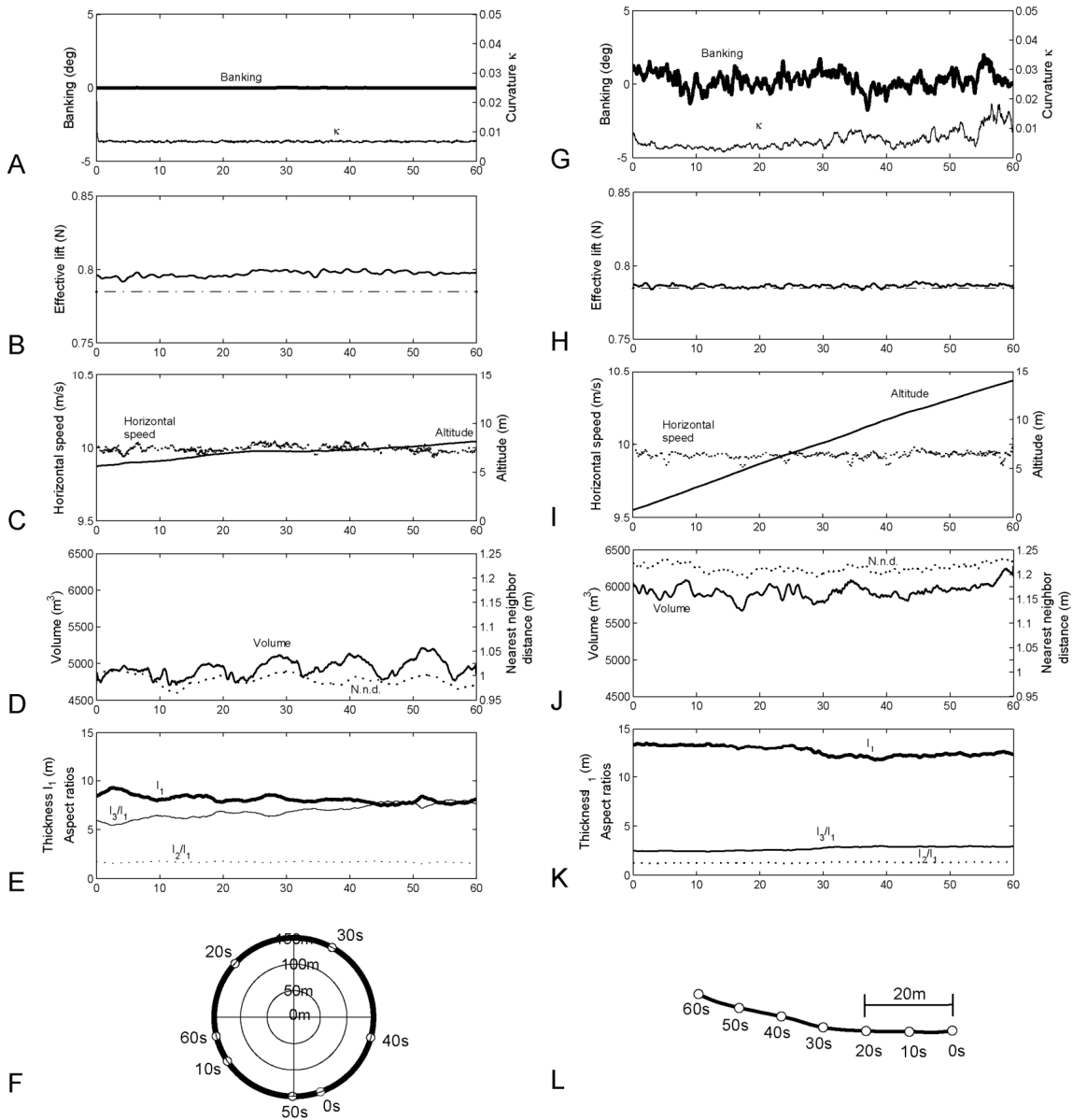

**Figure 10.** Time series (in seconds) of flight behaviour and flock properties in the model with attraction to the roost, drag and lift but without banking (at the left) and without attraction to the roost and with drag and lift and banking (at the right). A,G) average bank angle and path curvature, $\kappa$ ; B, H) Effective lift, dashed line represents lift at cruise speed and at level flight; C, I) horizontal speed and altitude; Note that without roost (I) altitude above the ground increases due to lack of vertical attraction to the roost, D, J) Volume and nearest neighbour distance (NND); E, K) Aspect ratio measured as $I_3{:}I_1$ (largest : shortest extent), $I_2{:}I_1$ (middle : shortest extent) and thickness ($I_1$). F.L) trajectories (spatial top view). Numbers indicate points in time. Flock size N=2000, default parameters.

The density is lower at the border than in the body of the flock despite the stronger attraction among individuals at the border (Fig. 7B, left side). Omission of the stronger attraction among individuals at the border worsen results significantly in the degree to which the density is higher in the center than at the border (in terms of NND) (Fig 7B, right side, Wilcoxon matched pairs signed ranks test comparing the differences in NND between border and interior with and without correction, N=60, t= 5550, P~0.00).

The low frequency of nearest neighbours ahead and behind depends on repulsion having no blind angle. If individuals in the model have a blind angle in the back of their separation zone, their nearest neighbours are usually located precisely behind and ahead and their distribution is no longer correlated with empirical data (Kendall tau, N=9, $\tau$=-0.11, NS). This arises because here, behind and ahead, they are located in each others' blind angle. Thus, they experience a weakened repulsion (compare the average angles of nearest neighbours in the model without blind angle (in blue squares) against those with blind angle (in white triangles), Fig. 7C).

**Discussion**

Our model shows complex patterns of flocking, such as the merging and splitting of flocks, and variability in shape and density of the flock. It shows close resemblance to both qualitative empirical data and to more than half of the aspects measured quantitatively. The resemblance to qualitative empirical data concerns the shape of flocks, their density, movement trajectories, turning behaviour, the tilting of the complete flock, splitting, merging and dynamics over time. The quantitative similarity to empirical data comprises aspects investigated in 10 statistical tests, i.e. shape (as regards two aspect ratios), the independence of one aspect ratio on the number of individuals and the volume of the flock, a significant correlation between thickness and volume and between volume and group size, the similarity of the density distribution at the front and the back of the flock, the independence of density from group size, the right skew of the nearest neighbour distribution and the orthogonal angle between shortest dimension (the thickness) and the flight direction of the flock, and the angles to the nearest neighbours. Eight tests do not match empirical data precisely. In the model shape (measured as $I_2/I_1$) depends on group size and volume, the volume of the flock and its (correlated) thickness, as well as its difference in density at the border versus the interior of the flock appear to be smaller than empirical estimates and modelled flocks stay too much parallel to the ground and at the same altitude above it.. .

In line with earlier models of swarming (Buhl et al., 2006; Couzin et al., 2002; Hemelrijk and Hildenbrandt, 2008; Hemelrijk and Kunz, 2005; Parrish and Viscido, 2005), our model shows that these displays of starlings may result from local interactions only. The interactions are local because the volume of interaction is approximately 1m$^3$ (Fig. 4) and the volume of the flock ranges between 500 and 17.000 m$^3$ (Tab. 2). It shows that neither global perception nor leadership is needed and only little cognition. Herewith, the model reduces perceptual and cognitive assumptions for the individual birds. As long as there is no evidence to the contrary, following Occam's razor, a model based on local perception and simple cognition is to be preferred to a model that assumes that individuals interact with the whole group.

As regards the causes of these patterns in the model, by taking traits out we have shown that all of the following traits are required to produce resemblance to the many patterns of aerial displays, namely banking (Fig. 10), attraction to the roost (Fig. 10), the hard sphere (Fig. 7A), the large flock size (Fig. 9) and all-round repulsion, i.e. the absence of the blind angle for repulsion at the back (Fig. 7C). Further, the variation in traits of the flocks over time (such as in vertical movement, density, volume and aspect ratios) is due to the combination of the large flock size, banking and the attraction to the roosting area, which induces turning. Through banking, turning leads to vertical movement, to variation in horizontal velocity and to a temporary increase in density of the flock and herewith to variation in volume and aspect ratios (Fig. 8).

The model seems to capture the essentials of aerial displays of starlings at the urban roost of Rome, in that a multitude of patterns in the model qualitatively resemble these displays. Further, because the flight forces are derived from the first principles of aerodynamic theory, because banking while turning is typical of birds and because the attraction to the roosting area is derived from empirical data, our model is useful to derive explanations for specific aspects of these displays and to function as a guide for empirical studies, as follows.

From the empirical data on flocking events in Rome, Ballerini and co-authors (Ballerini et al., 2008b) describe three traits, for which we develop through our model a hypothesis about their interconnection. The traits are the following: 1) while turning the shape of the flock changes relative to the direction of movement, 2) while turning all individuals of a flock follow paths of the same length with equal radii in real starlings, which rock doves do also (Pomeroy and Heppner, 1992) and 3) in starlings the neighbours that are nearest are seldom located behind or ahead. We confirm these three traits in our model (Fig. 6A, 7C). We speculate that they are interconnected as follows. Because individuals follow paths of the same length while turning, this means that individuals turn around their forward axis without speeding up or slowing down as would be required if they would take parallel paths. Thus their speed variance is low. It follows that in a turn the shape of the flock remains intact, but that its orientation in relation to the movement direction has changed. We speculate that the low variance of their speed and their high tendency to turn also has an effect on the angle under which individuals have their nearest neighbour (Fig. 7C). The explanation is as follows. In our model, we obtained a distribution of angles similar to empirical data by omitting the blind angle for separation at the back. Omission of the blind angle causes the individuals to avoid collisions by turning away more often and by speeding up or slowing down less frequently. This probably causes individuals to have their closest neighbours at their side more often. That real starlings turn easily is enabled by their short and broad wings (with a low aspect ratio) and the adaptive reason may be that starlings move fast and collision is dangerous. We further speculate that collisions caused by speeding up and slowing down may be more harmful than those that happen at an equal speed from the side during turning away. In sum, we think that the change of shape of the flock relative to the movement direction during turns, the equal turn radii and the lack of the neighbours along the movement direction may be side-effects of the low variance of speed and high tendency to turn that starlings aim for. More research is needed into these interrelationships. We are currently studying this hypothesis in a comparative study of our models of birds and fish (Hemelrijk and Hildenbrandt, 2008; Hemelrijk and Kunz, 2005; Kunz and Hemelrijk, 2003).

Like in real starlings the shape of the flock in our model is seldom oblong in the movement direction and thus, it differs from that of fish schools, which are often oblong. Nevertheless, the process that leads to an oblong school in fish operates in our model also, but the oblong shape is destroyed during turns. The main process that leads to an oblong shape in our former models of fish (Hemelrijk and Hildenbrandt, 2008;

Hemelrijk and Kunz, 2005; Kunz and Hemelrijk, 2003), is the avoidance of collision during coordinated movement. Individuals in our models of fish schools may try to avoid collisions by slowing down (Hemelrijk and Hildenbrandt, 2008) and thus increasing the length of the school or by turning away (Hemelrijk and Kunz, 2005; Kunz and Hemelrijk, 2003). Because this leaves a gap ahead of the individual that slowed down, the former neighbours move inwards, and thus, the school also becomes thinner. As a result the school becomes oblong. This process also works in our starling model, where individuals turn away rather than slowing down in the movement direction. It is visible most clearly if we switch off banking (Fig. 10D) or follow flocks in the complete model which do not turn for a prolonged period, for instance, when they cross the roost. The process of lengthening in our model of starlings is disturbed during sharp turns as becomes clear when we forbid sharp turns by omitting attraction to the roost and omitting banking (Fig. 10)..

In our complete starling model we observe not only a change in the shape relative to the movement direction in small flocks, but the relative proportions (aspect ratios) of the flock also change with flock size and the variation in shape is greater in large flocks than in small ones. The phenomena of variable aspect ratios over time and their dependence on group size differ from the findings by Ballerini et al (2008b). These authors stated that the aspect ratios of starling flocks were fixed and independent of flock size. However, our video recordings and qualitative observations of real starlings confirm that the shape of flocks is variable over time and influenced by group size. In our model this is a consequence of abrupt turns, which arise due to banking and due to attraction to the roost, and lead to the compression of the flock (Fig. 8). Because compression is smaller if the number of individuals is fewer, this variability of volume and shape is less pronounced in smaller flocks (Fig. 10). The conclusion of Ballerini and co-authors of fixed aspect ratios may be due to a bias in their selection of flock events. They may have only selected large (with more than 447 individuals) flocks that traveled forwards (i.e. were not turning). This is also what we needed to do to collect flocks in the model that matched their empirical data shown in table 2. To verify our model based suggestions, it should be measured in empirical data 1) whether aspect ratios of flock shape change over time, 2) whether turns are the major cause of these shape changes (in the absence of predation) and 3) whether aspect ratios differ between large and small flocks of real starlings.

The distributions of nearest neighbours in the empirical data of the two starling flocks are skewed to the right and there is a steep change around the length of their wingspan. This is explained by suggesting that individuals cannot overlap at closer distance - a so-called 'hard sphere' (Ballerini et al., 2008b). We confirm this hypothesis in our model, because if individuals are repulsed by a hard sphere, the right skew of the empirical distribution is reproduced, whereas without hard sphere repulsion, the model produces a distribution skewed to the left.

Density is independent of flock size both in the model and in empirical data of starlings (Ballerini et al., 2008b). This differs from findings in schools of fish, where density increases with group size (Breder, 1954; Keenleyside, 1955; Nursall, 1973; Partridge, 1980; Partridge et al., 1980). Our fish models suggest that this increase is due to the stronger mutual attraction between a higher number of individuals in larger schools (Hemelrijk and Hildenbrandt, 2008; Hemelrijk and Kunz, 2005; Kunz and Hemelrijk, 2003). The independence between density and flock size in starlings may have two causes. First, it may follow from the frequent turns of the flocks, because in turns groups are compressed and this effect on density may overshadow that of group size. Second, it may also be a consequence of the low and constant number of (topological) interaction partners in starlings (Ballerini et al., 2008a). This may prevent the mutual attraction from increasing in larger groups. However, that does not explain the positive correlations between density and flock size that have been found for starlings foraging on the ground (Williamson and Gray, 1975) and for rooks (Patterson, Dunnet, and Fordham, 1971). Thus clearly further studies are needed of the relationship between flock size and density. To omit the potential effects turning may have on the relation between density and flock size, density should be studied in empirical data of flocks of different sizes travelling outside the roost and in our starling model without attraction to the roost.

In small flocks in our model the banking of the flock is correlated with the banking of individuals during a turn (Fig 6B). This happens also in the empirical data: the complete flock tilts sideward in the direction of banking. For real starlings Ballerini and co-authors (Ballerini et al., 2008b) suggest that this may happen because individuals tend to keep their neighbours on the same visual plane to preserve information about their positions. However, in our starling model, individuals do not try to do so and still the flocks bank. A detailed study of flock banking in the model is needed to understand its underlying processes. This understanding may then be used as a hypothesis for an empirical study.

Flocks in Rome are flat, which is attributed to the fact that, due to gravity, vertical movement requires more energy than level flight (Ballerini et al., 2008b). However, in our model individuals also move preferentially at the same level, but this does not lead to flat flocks. In the model, flocks become flat due to the vertical attraction to a preferred height above the roost, arbitrarily named zero level. Since our way to induce a flock to become flat is arbitrary, other ways to obtain a flat flock should also be studied in the model.

Our model did not reproduce the empirical data as regards four phenomena, namely, the great volume and thickness of the starling flocks, their variable altitude above the ground and weakly parallel direction to it and their higher density at the border than the interior (Ballerini et al., 2008b). These differences may be interrelated and due to the absence in the model of disturbances such as threat of predation or interaction with other birds that induce a flock to turn. In the model the sharp turning that is induced by the border of the roost leads to a more variable altitude above the ground (Fig. 8C), and to an increase of the volume of the flock in general (compare Fig. 8D versus 10DJ) (and less strongly of its thickness), which may be due to a reshuffling of individuals as a consequence of the compression and expansion during sharp turns (Fig. 8D). This effect is stronger in a larger compared to a smaller flock, because the disturbance is stronger in a larger group (Fig. 9BC). The relevance of this hypothesis for real starlings may be investigated by studying whether flocks with the same number of individuals have a larger volume on average in the presence of a predator than without it. As to the lack of a higher density at the border than in the interior of the flock in the model despite our addition of stronger cohesion at the border (Hamilton, 1971), it may be that in nature attraction at the border due to predator threat is even stronger and that a recent evasive reaction may result in higher density at the border particularly at the side at which the predator was approaching. The low values for these four traits compared to empirical data may therefore be, among other things, due to the low rate of 'disturbance' of the flock in the model, whereas in reality flocks are disturbed continually, not only by attraction to the roost, but also, for instance, by the threat of predators (such as falcons) and of other species (such as gulls), the wind, rain and change in lighting conditions. Whether predator evasion (and reactions to other disturbing factors) increases the values of these traits of flocks in the model as well as in reality, will be one of the things to study in future.

In future this model may be extended to study effets of details of flying behaviour (such as flapping), foraging behaviour and evasion of attacks by predators. These may lead to extra variability in behaviour as mentioned above and observed in nature. For instance, as to predation, after extending our model with predators and comparing its patterns of evasion from predators to empirical data (Carere et al., 2009), we may also find the wave-like patterns that starlings produce when avoiding attacks by predators (Procaccini et al in prep). In starlings it is unclear whether these waves reflect changes in density, orientation or both as has been described for patterns of wave-like evasion in fish (Gerlotto et al., 2006), insects (Kastberger et al., 2008; Treherne and Foster, 1981) and humans (Farkas and Vicsek, 2006). Our future model may be helpful to decide this issue for starlings.

In its present form our model may be used as a general framework for the study of displays at the roost. This may be easiest as regards a comparison between flocks of starlings displaying at different roosts (Beauchamp, 1999; Brodie, 1976; Feare, 1984). For instance, the patterns observed at the roost in Termini were qualitatively similar to those observed at another wintering roost in the south of the city (EUR) (Carere et al., 2009),, but there may be quantitative differences that are of interest to study with the model. Similar patterns have also been observed for wintering roosts in other cities but quantitative data are not available either (Carere et al., 2009; Clergeau, 1990; Eastwood, Isted, and Rider, 1962; Feare, 1984; Williamson and Gray, 1975). Furthermore, we expect that with a few extensions, our model may also be used to investigate aerial displays at the roost by other species such as corvids (Danchin and Richner, 2001;

Gadgil, 2008 ), harriers (Gurr, 2008 ), shorebirds (Handel and Gill, 1992) and swallows (Russell et al., 1999 ; Smiddy et al., 2007).

Further, in our model robustness against parameter changes needs to be studied in more depth. Since we were interested in modeling the starling displays in Rome, we have not yet tried out parameters systematically (such as those of the size f the roost) apart from a range of lift/drag ratios (of up to 10), that were indicated in the literature (Möller, 2005). The results of the model remained similar for these ratios. Despite its many parameters, it should be noted that our model is an extreme simplification of reality.

We note that statistical testing of models is problematic in general (Roberts, 2000). In our model this is problematic also, since in it part of the parameters (of which no empirical data were available) have been tuned so as to make patterns in the model resemble flocking patterns of the video-recordings of real starlings. However, whereas we tune for instance the average nearest neighbour distance of a flock, we study traits which we did not tune, such as the distribution of distances and the angles to the nearest neighbours. Statistical tests should be seen as a quantitative measure of it's the model's similarity to empirical data. This degree of similarity is important for the usefulness of our model to get insight in empirical data and guide future empirical studies.

A general criticism of models with many parameters that is often heard is that they lack explanatory value, because they can be made to fit single patterns too easily. However, in our model, we have generated most patterns as emergent phenomena from the combination of coordination, simplified flying behaviour, and attraction to the roost. Here, we follow the so-called bottom-up procedure of tracing the effects of simple behavioural rules on collective patterns (Hemelrijk, 2002). However, two of the patterns we generated did not emerge in this way. This is the case for the flatness of the flock (which we induced by vertical attraction to a preferred height above the roost), and for the angle under which individuals have their nearest neighbours (for this we drove close by neighbours out of the blind angle by closing the blind angle). Here, we searched for what needed to be changed in the model to generate these patterns, thus, we used pattern-oriented modeling (Grimm et al., 2005). This is useful too, because these two adaptations led to hypotheses that can be studied further (see above). Further, as to the dynamics over time, quantitative empirical data are lacking, but the patterns of the model resemble those of rock doves (Pomeroy and Heppner, 1992) and starlings qualitatively (Fig. 8). Here, our model may be used as a 'prediction' for empirical data. In general, since studies of huge swarms of starlings are labour intensive, our model-based hypotheses may be useful in indicating suitable topics for empirical study.

In sum, we demonstrate in a model that local interactions and self-organisation suffice to reproduce patterns of aerial displays of starlings qualitatively and in many cases also quantitatively, but only if they are combined with specific aspects of the flying behaviour and the environment (i.e. attraction to a roosting area). It herewith provides us with a suitable tool for further study of these collective displays.

**Acknowledgement**

We would like to thank Andrea Cavagna, Volker Grimm and Daan Reid for comments on a former draft. We are grateful to Dirk Visser for drawing figures 1, 2 and 5a. We thank Daniela Santucci, Enrico Alleva for criticism and discussions and Francesca Zoratto for technical support in the field. This study was partly financed by a grant from the 6$^{th}$ European framework under the STREP-project 'StarFlag' (n12682) in the NEST-programme of 'Tackling complexity in science'.

Self-organised starling displays, 44## References